\documentclass[12pt]{iopart}

\newcommand{\beq}{\begin{equation}}
\newcommand{\enq}{\end{equation}}
\newcommand{\ket}[1]{|{#1}\rangle}
\newcommand{\bra}[1]{\langle {#1}|}
\begin{document}
\title{Collective excitations in an $F=2$ Bose-Einstein condensate}
\author{J.-P. Martikainen$^\dagger$ and K.-A. Suominen$^{\dagger,\ddagger}$}

\address{$^\dagger$Helsinki Institute of Physics, PL 64, FIN-00014 Helsingin
yliopisto, Finland\\
$^\ddagger$Department of Applied Physics, University of Turku, FIN-20014 
Turun yliopisto,
Finland}


\begin{abstract}
We calculate the collective excitations in a homogeneous $F=2$ spinor 
condensate in
the absence of the magnetic field and also in a strong magnetic 
field. Almost all the
excitations for a zero magnetic field are found to have either the 
Bogoliubov form
or the free-particle form. We relate our results for a strong 
magnetic field to the
concept of fragmented condensates and note that some observable 
quantities such as
the speed of sound for the cyclic state depend on the existence 
of relative phase between the spinor amplitudes for the $M=\pm 2 ,0$ atoms.
\end{abstract}
\pacs{03.75.Fi, 32.80.Pj, 03.65.-w}

\maketitle    
\section{Introduction}\label{Intro}

Due to experimental observations of Bose-Einstein condensation in dilute alkali
gases~\cite{Anderson95,Bradley95,Davis95} the weakly interacting 
quantum systems
have become a highly relevant research topic. The spectrum of collective
excitations for such a Bose condensate was predicted some time ago by
Bogoliubov~\cite{Bogoliubov} and since then these studies have been extended to
trapped systems in a number of papers, see 
reference~\cite{Dalfovo_review}. Recently a
sodium condensate of $^{23}{\rm Na}$ atoms in the $F=1$ hyperfine 
state was trapped
in an optical dipole trap~\cite{Stamper98a}. Such a trap confines 
simultaneously all
the different Zeeman substates (quantum number $M$) of the atomic 
hyperfine ground
state  states (quantum number $F$), whether or not their energy degeneracy is
removed by a magnetic field. Therefore the spin degree of freedom is 
not necessarily
frozen, and the condensate must be described in terms of a 
multicomponent spinor
wavefunction.

The possible ground states and excitations of the $F=1$ spinor 
condensates have been
studied theoretically by several 
authors~\cite{Ho98,Ohmi98,Isoshima99}. So far only
a few studies have been made for the $F=2$ condensates, such as the 
calculation of
the possible ground states, and their dependence on a weak magnetic 
field by Ciobanu
et al.~\cite{Ciobanu00}. One can identify several possible ground 
states, and the
relative magnitude of the scattering lengths for different $M,M'$ 
collisions ($M+M'$
is conserved) determines which of them is the true ground state. Our 
results for the
ground states agree with those presented in reference~\cite{Ciobanu00}, 
and we calculate
further the collective excitations associated with each possible ground state.

Condensates have been produced on an $F=2$ state experimentally in
magnetic trap for
$^{87}$Rb~\cite{Arlt99,Mueller00}. Also, one can achieve condensation 
on an $F=1$
state, and then move the atoms with suitable microwave pulses to the 
$F=2$ hyperfine
ground state~\cite{Myatt97}. In that case the magnetic trap potentials are not
affected by the change in the internal state of the atoms, so one 
could imagine of
producing a condensed sample of atoms in a magnetic trap on the $F=2$ 
state, and
then moving it into an optical trap, after which one can remove the 
magnetic field
and thus return to the case of degenerate Zeeman states. The main 
problem in this
case are the inelastic atomic collisions, which may transfer atoms 
from the $F=2$
manifold into the $F=1$ manifold. However, for $^{87}$Rb these rates 
are surprisingly
small~\cite{Burke97,Suominen98}. Also, this problem is not present e.g. for
$^{85}$Rb as then the $F=2$ hyperfine state is the lowest ground 
state. Very recently~\cite{Barrett2001a} a condensate of $^{87}$Rb atoms
on the $F=1$ state was obtained in an all-optical trap via evaporative
cooling, and this approach is expected to work also for atoms in the
$F=2$ state.

This paper proceeds as follows. In section~\ref{Model} 
we introduce the physical
model for the interacting spinor system, and in 
section~\ref{Pos_Ground_states} we
discuss the ground states of the $F=2$ condensate. We calculate the collective
excitations in the absence of magnetic fields in 
section~\ref{Excitations}. We also
study how the excitations change if a strong magnetic field is applied
(section~\ref{Big_B}) and connect the results to the concept of 
fragmentation in spinor
condensates (section~\ref{fragmented_state}). 
We also briefly note the superfluid
properties of the different possible ground states in 
section~\ref{superfluid}, and
discuss our results in the concluding section~\ref{conclusion}.

\section{Physical model}\label{Model}

The weak interaction between condensed atoms is generally described 
very well with
a contact potential whose strength is proportional to the $s$-wave scattering
length $a$.  This has been thoroughly demonstrated with successful comparison
of theory with experiments on single-component condensates and also for
two-component condensates, see e.g. reference~\cite{Leggett01} and 
references therein. We
generalize this interaction potential to apply for the $F=2$ spinor 
condensates and
use the interaction potential~\cite{Pu99}
\beq
   	V(r_2-r_1)=\delta(r_2-r_1)\sum_{f=0}^4g_f\sum_{m=-f}^f
   	\ket{f,m}\bra{f,m},\label{interaction_pot}
\enq
where $\ket{f,m}$ is the total (molecular) hyperfine spin state formed by two
atoms with spin $F=2$, and $g_f=4\pi\hbar^2 a_f/m_a$ with $a_f$ being the
scattering length in the $f$ channel and $m_a$ the atomic mass.

Following the formalism of Pu {\it et al.} for the $F=1$ 
condensate~\cite{Pu99}, we
expand this interaction term in terms of the two-atom basis vectors
$\ket{F_1=2,M_1} \otimes \ket{F_2=2,M_2} = \hat{\Psi}_{M_1} \otimes
\hat{\Psi}_{M_2}$. This procedure is straightforward, but the 
resulting equation is
very long and thus we omit it here. To simplify things we limit our 
discussion to a
{\it homogeneous} condensate and therefore the total Hamiltonian 
includes only the
interaction  potential and the kinetic energy $K=-[\hbar^2/(2m_a)]\nabla^2$. We
also make the semiclassical approximation and replace the atomic creation and
annihilation operators with complex numbers $\Phi_{M}$.

The interaction term contains terms proportional to densities 
$n_M=|\Phi_{M}|^2$ of
the different $M$ components, but also terms that could result in complex
spin-mixing dynamics, {\it i.e.} transient behavior of 
the populations in different $M$ states.
In the $F=2$ case these processes are expected 
to be much more complicated than in the $F=1$ condensate studied by Pu 
{\it et al.}~\cite{Pu99}. To give an idea about the resulting 
Gross-Pitaevskii~\cite{Pitaevskii1961a,Gross1961a} (GP) 
equations we present the equation for the $M=0$ component:
\begin{eqnarray}
\hspace{-1cm}i\hbar\frac{\partial \Phi_0}{\partial t}&=&\left[K+
\lambda_s\sum_{M=-2}^2 n_M
+\alpha\left(n_1+n_{-1}\right)+\beta\left(n_2+n_{-2}\right)\right]\Phi_0
+2\alpha\Phi_0^*\Phi_1\Phi_{-1}\nonumber\\
&&-2\beta\Phi_0^*\Phi_2\Phi_{-2}+\gamma\left(
\Phi_{-1}^*\Phi_{-2}\Phi_{1}+\Phi_{1}^*\Phi_{2}\Phi_{-1}
+\Phi_{-2}^*\Phi_{-1}^2+\Phi_{2}^*\Phi_{1}^2\right),
\end{eqnarray}
where $\lambda_s=(18g_4+10g_2+7g_0)/35$, $\alpha=(12g_4-5g_2-7g_0)/35$, 
$\beta=(-3g_4+10g_2-7g_0)/35$, and $\gamma=\sqrt{6}\,(g_4-g_2)/7$.

We mostly ignore mixing dynamics and focus on the condensate
behavior around stationary states (ground states).  At first we assume that the
external magnetic field is absent, but in section~\ref{Big_B} we discuss the
behavior of the stationary states in a strong magnetic field.

\section{Possible ground states}\label{Pos_Ground_states}

Ground state structures for a $F=2$ spinor condensate were calculated 
by Ciobanu
{\it et al.}~\cite{Ciobanu00}. We follow their notation and terminology for the
various ground states ($F,F',C,P,P0$ and $P1$). The ground states and 
their energies
are given in table~\ref{Ground_states} for the case when the magnetic field is
absent. When comparing these results with results in 
reference~\cite{Ciobanu00} it should
be noted that in reference~\cite{Ciobanu00} the energy of the cyclic state 
was chosen to be zero. Taking this shift into account the results in 
table~\ref{Ground_states} coincide with those in reference~\cite{Ciobanu00}.

\begin{table}
\begin{center}
\caption[Table1]{The possible ground states (spinor amplitudes 
$\overline{\Phi}$ and energies $E$) for
the $F=2$ condensate in zero magnetic field. The angle variables
$\phi,\alpha_{\pm 1}$ and $\alpha_{\pm 2}$ can be chosen freely.
\label{Ground_states}}
\begin{tabular}{ccc}
&
\( \overline{\Phi } \)&
\( E \)\\
\hline
F&
\( (1,0,0,0,0) \)&
\( ng_{4} \)\\
F'&
\( (0,1,0,0,0) \)&
\( n(4g_{4}+3g_{2})/7 \)\\
C&
\( \frac{1}{2}\left( e^{i\phi },0,\sqrt{2},0,-e^{-i\phi }\right)  \)&
\(n(3g_{4}+4g_{2})/7 \)\\
P&
\( \frac{1}{\sqrt{2}}\left( e^{i\alpha _{2}},0,0,0,e^{i\alpha 
_{-2}}\right)  \)&
\( n(18g_{4}+10g_{2}+7g_{0})/35 \)\\
P1&
\( \frac{1}{\sqrt{2}}\left( 0,e^{i\alpha _{1}},0,e^{i\alpha 
_{-1}},0\right)  \)&
\( n(18g_{4}+10g_{2}+7g_{0})/35 \)\\
P0&
\( \left( 0,0,1,0,0\right)  \)&
\( n(18g_{4}+10g_{2}+7g_{0})/35 \)\\
\end{tabular}
\end{center}
\end{table}

Within the semiclassical approximation the true ground state is the state
with the lowest energy in table~\ref{Ground_states}. This 
approximation may ignore
some subtle effects. For example, in the $F=1$ spinor condensate with
an antiferromagnetic interaction  the true ground state will be a fragmented
condensate~\cite{Ho00}. In such a condensate the number of particles in each
component is separately constant and there is no well defined 
phase relationship between different components.
How much this matters in practice is still a matter
of debate~\cite{Javanainen00}.
The  semiclassically calculated results capture only
states corresponding to a single condensates. In the $F=1$ spinor 
condensate the
energy difference  between single and fragmented condensates vanishes at the
thermodynamic limit so we expect that most energies in
table~\ref{Ground_states} will be correct (in the thermodynamic limit) even if
the true ground state would be fragmented. The cyclic state~\cite{Klausen01} 
is a possible exception. We  will return to this issue in
section~\ref{fragmented_state}.

	As polar states are degenerate one might assume that a 
superposition of polar states would always give a state with same energy. This
is clearly wrong. The cyclic state is a special superposition of
$P0$ and $P$ states, but it has nevertheless a different energy. 
Forming a superposition
of polar states will change the systems energy non-linearly. In particular,
the Hamiltonian has a term 
\begin{eqnarray}
	H_{\pm 2,0}&=&\frac{-3g_4+10g_2-7g_0}{35}\left[
\Phi_{-2}^*\Phi_{0}^*\Phi_{-2}\Phi_{0}+\Phi_{2}^*\Phi_{0}^*\Phi_{2}\Phi_{0}
\right.\nonumber\\
&&\left.-\Phi_{0}^*\Phi_{0}^*\Phi_{-2}\Phi_{2}-
\Phi_{-2}^*\Phi_{2}^*\Phi_{0}\Phi_{0}\right]
\end{eqnarray}
which does not contribute to the $P0$ and $P$ states separately, but will
contribute to their superposition. For non-magnetic states and 
when $3g_4-10g_2+7g_0>0$ this term is minimized by having $1/4$ of the 
population
at the $M=2$ state and a relative phase $2\phi_0-\phi_2-\phi_{-2}=\pi$.
This term is responsible for the peculiar phase-locking of the
cyclic state. 

\section{Excitations in zero magnetic field}\label{Excitations}

In our approach we simply linearize the Gross-Pitaevskii (GP) 
equations for each
component around all stationary states. The resulting equations  can
be solved to get the energies of the collective excitations. Many of these
equations are coupled, but all of them can be solved analytically. Some of the
coupled equations have the form studied previously by
Timmermans~\cite{Timmermans98}, and some lead to a $3\times 3$ 
eigenvalue problem.
In the following subsections and tables~\ref{Excitations_B0_Ferro},
\ref{Excitations_B0_Polar}, and~\ref{Excitations_B0_Cyclic} we 
present the results of our analysis. Our notation is such that
index numbers refer to the $M$ states present in the excitation and
index alphabets are used (when necessary) as labels to distinguish between
excitations involving the same $M$ states. A negative excitation energy implies
a thermodynamic (or energetic) instability since the system energy can be
lowered by creating such excitations. On the other hand, 
an imaginary value of the excitation energy implies dynamical instability.
This instability results in exponentially growing excitation amplitude
and can occur even in the absence of dissipation.

\subsection{Ferromagnetic states}

The results given in table~\ref{Excitations_B0_Ferro} show that the excitations
of the ferromagnetic state $F$ are either of the Bogoliubov form ({\it i.e.} 
$E=\sqrt{K(K+2\epsilon)}$) or the free-particle
form $E=K+\Delta E_g$ , where $\Delta E_g$ is the energy gap. The $F'$
state, on the other hand, seems more interesting. It is 
not stable energetically
since when the magnetized ground state is favored the $F$ state has 
always a lower energy, but experimental preparation could leave the 
condensate in the $F'$ state,
after which the relaxation to the $F$ state is slow and thus the $F'$ 
state could
appear as a metastable one. We have the Bogoliubov branch and two free particle
excitations but there are also two excitations which do not fall into these two
basic categories. These excitations have the form 
\beq
	E=-\frac{\Delta E}{2}\pm \frac{\sqrt{(\Delta E+2K)^2+16K\Delta E}}{2},
\enq
where $\Delta E=n(g_4-g_2)/7$. At small kinetic energies the other 
one approaches $0$ and the other one $-\Delta E$.

When $g_4-g_2>0$ all excitations are real and $F'$ is thermodynamically
unstable as excitations $E_{20-}$ and $E_{-2}$ both have negative values at
small values of kinetic energy. On the other hand, if $g_4-g_2<0$ the 
$F'$ state
is also dynamically unstable as the $E_{20\pm}$ modes have imaginary energies.
The imaginary value takes a maximum value when $K=5n(g_2-g_4)/14$.
Presumably the related value of the wavenumber $k$ sets the scale for
the spatial structures caused by the dynamical instability.

\begin{table}[bt]
\begin{center}
\caption[Table2]{Excitations in zero magnetic field for the 
ferromagnetic states
$F$ and $F'$.  The subscripts in the names of the branches refer to the
$M$ state or states involved in the excitation superposition.
\label{Excitations_B0_Ferro} }
\begin{tabular}{l}
\hspace{2.0cm}F\\
\hline
\( E_{2}=\sqrt{K(K+2ng_{4})} \)\\
\( E_{1}=K \)\\
\( E_{0}=K-\frac{4}{7}n\left( g_{4}-g_{2}\right)  \)\\
\( E_{-1}=K-\frac{6}{7}n\left( g_{4}-g_{2}\right)  \)\\
\( E_{-2}=K+\frac{2}{35}n\left( -17g_{4}+10g_{2}+7g_{0}\right)  \)\\
\hline  
\hspace{4.5cm}F'\\
\hline
\( E_{2,0,a}=\frac{1}{2}\left[ -\frac{n(g_{4}-g_{2})}{7}+\sqrt{\left(
\frac{n(g_{4}-g_{2})}{7}+2K\right) 
^{2}+\frac{16}{7}Kn(g_{4}-g_{2})}\right]  \)\\
\( E_{2,0,b}=\frac{1}{2}\left[ -\frac{n(g_{4}-g_{2})}{7}-\sqrt{\left(
\frac{n(g_{4}-g_{2})}{7}+2K\right) 
^{2}+\frac{16}{7}Kn(g_{4}-g_{2})}\right]  \)\\
\( E_{1}=\sqrt{K\left[ K+\frac{2}{7}n\left( 4g_{4}+3g_{2}\right) \right] } \)\\
\( E_{-1}=K+\frac{2}{35}n\left( -2g_{4}-5g_{2}+7g_{0}\right)  \)\\
\( E_{-2}=K-\frac{3}{7}n\left( g_{4}-g_{2}\right)  \)\\
\end{tabular}
\end{center}
\end{table}

\subsection{Polar states}

Table~\ref{Excitations_B0_Polar} shows that all excitations in the polar states
have the Bogoliubov form. As the chemical potentials of these states are the
same, the main branch is identical in all polar states. If the $P$ 
state is to be
dynamically stable, then the terms $x=17g_4-10g_2-7g_0$ and $y=-3g_4 
+10g_2 -7g_0$
must be positive.  This in turn implies that also the $P0$ state would be
dynamically stable. On the other hand, the $P1$ state can be 
dynamically unstable if
$y<\frac{x}{3}$ or $y>3x$. This instability is due to the imaginary energies in
either of the modes $E_{\pm 2,0,b}$ or $E_{\pm 2,0,c}$ and will be 
reflected in the formation of domains ({\it i.e.} stripes) 
with $M=\pm 2,0$ atoms. 


\begin{table}[bt]
\begin{center}
\caption[Table3]{Excitations in zero magnetic field for polar states.   The
subscripts in the names of the branches refer to the
$M$ state or states involved in the excitation superposition.
\label{Excitations_B0_Polar} }
\begin{tabular}{l}
\hspace{3.7cm}P\\
\hline
\( E_{\pm 2,a}=\sqrt{K\left[ K+\frac{2n}{35}\left( 
18g_{4}+10g_{2}+7g_{0}\right)
\right] } \) \\
\( E_{\pm 2,b}=\sqrt{K\left[ K+\frac{2n}{35}\left( 
17g_{4}-10g_{2}-7g_{0}\right)
\right] } \)\\
\( E_{0}=\sqrt{K\left[ K+\frac{2n}{35}\left( 
-3g_{4}+10g_{2}-7g_{0}\right) \right]
} \)\\
\( E_{\pm 1}=\sqrt{K\left[ K+\frac{2n}{35}\left( 2g_{4}+5g_{2}-7g_{0}\right)
\right] } \)\\
\hline
\hspace{3.7cm}P1\\
\hline
\( E_{\pm 1,a}=\sqrt{K\left[ K+\frac{2n}{35}\left(
18g_{4}+10g_{2}+7g_{0}\right)\right] } \)\\
\( E_{\pm 1,b}=\sqrt{K\left[ K+\frac{2n}{35}\left( 2g_{4}+5g_{2}-7g_{0}\right)
\right] } \)\\
\( E_{\pm 2,0,a}=\sqrt{K\left[ K+\frac{2n}{35}\left( 
2g_{4}+5g_{2}-7g_{0}\right)
\right] } \)\\
\( E_{\pm 2,0,b}=\sqrt{K\left[ K+\frac{2n}{35}\left( 
-13g_{4}+20g_{2}-7g_{0} \right)
\right] } \)\\
\( E_{\pm 2,0,c}=\sqrt{K\left[ K+\frac{2n}{35}\left( 
27g_{4}-20g_{2}-7g_{0}\right)
\right] } \)\\
\hline  
\hspace{3.7cm}P0\\
\hline
\( E_{0}=\sqrt{K\left[ K+\frac{2n}{35}\left( 
18g_{4}+10g_{2}+7g_{0}\right)  \right]
} \)\\
\( E_{\pm 2}=\sqrt{K\left[ K+\frac{2n}{35}\left( -3g_{4}+10g_{2}-7g_{0}\right)
\right] } \)\\
\( E_{\pm 1}=\sqrt{K\left[ K+\frac{2n}{35}\left( 12g_{4}-5g_{2}-7g_{0}\right)
\right] } \)\\
\end{tabular}
\end{center}
\end{table}

\subsection{Cyclic state}

Most excitations in the cyclic state have the Bogoliubov form
(table~\ref{Excitations_B0_Cyclic}) with the exception the mode 
$E_{\pm 2,0,b}$,
which has the free particle form. This mode is a superposition of $M=\pm 2,0$
atoms of the type $\left(-1,0,\alpha,0,1\right)$. When the cyclic state is the
ground state, this excitation has a positive gap. When the polar 
state has a lower
energy this mode has a negative gap, indicating thermodynamical instability. 


\begin{table}[bt]
\begin{center}
\caption[Table4]{Excitations in zero magnetic field for the cyclic state.
The subscripts in the names of the branches refer to the
$M$ state or states involved in the excitation superposition.
\label{Excitations_B0_Cyclic} }
\setlength{\tabcolsep}{0mm}
\begin{tabular}{l}
\hspace{3.0cm}C\\
\hline
\( E_{\pm 1}=\sqrt{K\left[ K+\frac{4n}{7}\left( g_{4}-g_{2}\right) 
\right] }  \)\\
\( E_{\pm 2,0,a}=\sqrt{K\left[ K+\frac{4n}{7}\left( 
g_{4}-g_{2}\right) \right] }
\)\\
\( E_{\pm 2,0,b}=K+\frac{2n}{35}\left( 3g_{4}-10g_{2}+7g_{0}\right)  \)\\
\( E_{\pm 2,0,c}=\sqrt{K\left[ K+\frac{2n}{7}\left( 
3g_{4}+4g_{2}\right) \right] }
\)\\
\end{tabular}
\end{center}
\end{table}

\section{Excitations in a strong magnetic field}\label{Big_B}

Let us assume that the condensate is prepared in one of the possible ground
states given in table~\ref{Ground_states}. Next we turn on a strong
longitudinal magnetic field ${\bf B}=B\hat{z}$, 
for which the quadratic Zeeman effect is at least 
on the order of the chemical potential. In a magnetic
field the true ground state will be something different, but if dissipative
processes are slow enough the condensate can still be metastable. Of 
course, the
scattering lengths can be strongly affected by the magnetic field, and new
inelastic two-body loss channels appear due to the Zeeman shifts, but 
we believe
it still worthwhile to briefly examine what might happen at such a situation.

In the presence of a magnetic field the different $M$ states are not 
degenerate, and
for strong fields the energy separation of adjacent states are not 
equal, either. The Zeeman shifts for the $M$ states are given by 
$E_{Z,M}=-\mu_B\left(Mg_FB+M^2g_F^{(2)}B^2\right)$, where $\mu_B$ is the
Bohr magneton, and $g_F$ and $g_F^{(2)}$ are the linear and quadratic 
Zeeman coefficients.
The terms in the Gross-Pitaevskii equations that are responsible 
for spin-mixing
dynamics begin to acquire time-dependent phase factors. As the 
magnetic field is
strong these phase factors vary on a timescale much shorter than 
everything else in
our system. This allows us to average over them, leaving the Gross-Pitaevskii
equations without the spin-mixing terms.

The chemical potentials for different components
are now different. We write the wavefunction as 
$\Phi_M(t)=\Psi_M\exp\left(-i\mu_M t/\hbar\right)$ and solve for the
chemical potential $\mu_M$.
For polar and ferromagnetic states they are given by
$\mu_M=\mu_0+E_{Z,M}$, where $\mu_0$ is the chemical potential
in the absence of magnetic field (see previous sections). 
For the cyclic state the the chemical potentials 
for $M=\pm 2,0$ atoms are $\mu_M=\frac{n}{70}(33g_4+30g_2+7g_0)+E_{Z,M}$, 
{\it i.e.}
the constant in front of the Zeeman shift is different from the zero field
chemical potential.

We have calculated the excitations without the spin-mixing terms in 
the same manner as before.  
The main branches of the collective excitations ({\it i.e.}, the 
ones having a form $\sqrt{K(K+2\mu_0)}$) are unchanged for
all the other states except for the cyclic state. For the cyclic state
the main branch is now $\sqrt{K\left(K+\frac{n}{35}(33g_4+30g_2+7g_0)\right)}$.
In the absence of
spin-mixing terms the components with very little population
follow an ordinary free particle Schr\"{o}dinger equation. In particular,
the terms proportional to the complex conjugate of the wavefunction are 
missing. Therefore excitations corresponding to these components 
take the free particle form with different gaps. 

In addition to the main branch the $P$ and
$P1$ states have also another Bogoliubov-type excitation with energies $E_{\pm
2,b}$ and $E_{\pm 1,b}$ respectively (here we follow the notation for the
zero field results). These Bogoliubov type excitations result from 
two component GP equations of the type studied by 
Timmermans~\cite{Timmermans98}. For the $F'$ state all excitation, except the
main branch, take the free particle form.
This implies that possible dynamical 
instabilities of the $F'$ state have been replaced with thermodynamical 
instabilities.

For the cyclic states we are left with three excitations of the Bogoliubov 
form, the main branch and two others (table~\ref{Excitations_BigB_Cyclic}). 
These involve atoms from the components with large populations.
The cyclic state can still 
be dynamically stable if $\left(17g_4-10g_2-7g_0\right)>0$.

As for the thermodynamical stability of the cyclic state 
we have three magnetization conserving channels available. The first 
channel (a) 
is to take atoms from $M=\pm 2$ states and
put them into $M=\pm 1$ states, the second channel (b) 
is to take two atoms from
the $M=0$ state and put them into $M=\pm 1$ states, and the last channel 
(c) is to
take two atoms from the $M=0$ state and put them into $M=\pm 2$ states.
The first and second processes tell us about relaxation towards the $P1$ state
and the third process indicates relaxation to the $P$ (or $P0$) state.
To ensure dynamical stability we will assume that  
$\left(17g_4-10g_2-7g_0\right)>0$.
The three processes come with energy costs
\begin{eqnarray}
	\Delta E_a&=&E_{required}-E_{released}=\nonumber\\&&
2K-6\mu_Bg_F^{(2)}B^2+\frac{n}{35}\left(17g_4-10g_2-7g_0\right)
\end{eqnarray}
\begin{eqnarray}
	\Delta E_b&=&E_{required}-E_{released}=\nonumber\\
&&2K+2\mu_Bg_F^{(2)}B^2+\frac{n}{35}\left(17g_4-10g_2-7g_0\right)
\end{eqnarray}
\beq
	\Delta E_c=E_{required}-E_{released}=2K+8\mu_Bg_F^{(2)}B^2.
\enq

If $g_F^{(2)}$ is positive and the term 
$\frac{n}{35}\left(17g_4-10g_2-7g_0\right)$ is
small (which is very likely) one expects the process (a) 
to dominate since this does not necessarily cost energy. This is reflected 
in the increase of the number of $M=\pm 1$ atoms and in the reduction of the
number of $M=\pm 2$ atoms. Finally, the spin-changing
processes will drive the system towards the true ground state, the $F$ state.
If, on the other hand,
the term $\frac{n}{35}\left(17g_4-10g_2-7g_0\right)$ turns out to be
larger than $6\mu_Bg_F^{(2)}B^2$, 
all three processes cost energy, forming an energy barrier, 
and the cyclic state can be metastable.

If $g_F^{(2)}$ is negative the process (c) does not cost energy. Also, if
$\frac{n}{35}\left(17g_4-10g_2-7g_0\right)$ is small
the process (b) does not cost energy either, but as there is more energy
to be gained in process (c) one expects a relaxation to the $P$ state, until
spin-changing processes start to dominate. 

If there is some other mechanism, beside the magnetic field, that
randomizes the phase one could apply the strong field results in the
limit $B\rightarrow 0$ and the metastability of the cyclic state
might be a more relevant issue.  

\begin{table}[bt]
\begin{center}
\caption[Table5]{Excitations in a strong magnetic field for the cyclic state.
The subscripts in the names of the branches refer to the
$M$ state or states involved in the excitation superposition. The energies 
for the $M=\pm 1$ atoms are scaled by $\frac{n}{70}(33g_4+30g_2+7g_0)$. 
\label{Excitations_BigB_Cyclic} }
\setlength{\tabcolsep}{0mm}
\begin{tabular}{l}
\hspace{3.7cm}C\\
\hline 
\( E_{\pm 2,0,a}=\sqrt{K\left[ K+\frac{1}{35}n\left( 33g_{4}+30g_{2}+7g_{0}\right) \right] } \)\\
\( E_{\pm 2,0,b}=\sqrt{K\left[ K+\frac{1}{35}n\left( 17g_{4}-10g_{2}-7g_{0}\right) \right] } \)\\
\( E_{\pm 2,0,c}=\sqrt{K\left[ K+\frac{1}{35}n\left( 3g_{4}-10g_{2}+7g_{0}\right) \right] } \)\\
\( E_{1}=K+E_{Z,1}+\frac{n}{70}\left( 17g_{4}-10g_{2}-7g_{0}\right)  \)\\
\( E_{-1}=K+E_{Z,-1}+\frac{n}{70}\left( 17g_{4}-10g_{2}-7g_{0}\right)\)\\
\end{tabular}
\end{center}
\end{table}

\section{Fragmented state}\label{fragmented_state}

In a fragmented condensate the numbers of particles in different $M$ states are
separately constant~\cite{Leggett01,Ho00} and the terms in the
Hamiltonian giving rise to the spin mixing dynamics average to zero. 
This suggests that we are describing the system
in the framework of fragmented states as soon as we have blocked
the spin-mixing dynamics.~
\footnote{
Often the precise nature of the condensate state is
more or less irrelevant. The measurement process can, for example, ``induce''
a relative phase even between number states~\cite{Javanainen96}. 
Also the density distribution
is accurately given by the GP equation even though its theoretical 
justification in the case of number states is dubious. Due to these 
complications the wave function in the absence of spin mixing should
be considered no more than the square root of the density. Whether the act of 
measurement can induce a well defined phase to the different components 
(and thus destroy the fragmented state) is an interesting problem that
should be addressed.
}

Thus our previous results for the $F=2$ spinor
condensate in a strong magnetic field could be equivalent to a study
of a spinor condensate in a framework of fragmented states. There is another
argument which supports this interpretation.  In a fragmented state 
the relative phase between different $M$ state  wavefunctions is not 
a meaningful concept. We
can consider the relative phases as random variables between $0$ and 
$2\pi$~\cite{Leggett01}.  As
this randomness is present for all timescales we can replace all the terms
which depend on a relative phase  of different $M$ states, with zero, their
average value.

Most of the chemical potentials do not change when spin-mixing terms are
dropped. As in the thermodynamic limit the energy of the $F=1$ single 
condensate
is the same as the energy  of the fragmented state, we find it quite 
likely that
an unchanged chemical potential reflects the connection between frameworks for
single condensates  and fragmented condensates. Especially so since the only
state whose chemical potential is 
affected by ignoring spin-mixing dynamics is the cyclic state, the only
fundamentally new state not already present in the $F=1$ condensate.

We conclude that many collective excitations for ferromagnetic and
polar states are different in a fragmented $F=2$ condensate, compared to the
zero field case. The main branches are not changed, but especially excitations
that are superpositions of different $M$ states are replaced with particle-like
excitations. Also many possible dynamical instabilities of the single 
condensate
are replaced with thermodynamic instabilities implying, perhaps, that the
relaxation rates to the true ground states are different for the fragmented
condensates  as opposed to the single condensates.

As the chemical potential of the cyclic state depends on whether we keep the
spin-mixing terms or not, it follows that the chemical  potential depends on
whether or not the relative phase of the $M=2,0,-2$ components is well defined.
Also, since some physical properties such as sound velocity, for 
example, depend
on the chemical potential we conjecture that if the relative phase in a cyclic
state is well defined (as it should be) there should be a change in the sound
velocity as we move from a zero magnetic field to a strong one. The
magnitude if this shift is given by
\begin{eqnarray}
\Delta c_s&=&c_{B\gg 0}-c_{B=0}\nonumber\\
&=&\sqrt{\frac {n}{7m}}\left(
\sqrt{\frac{33g_4+30g_2+7g_0}{10}}-\sqrt{3g_4+4g_2}\right).
\end{eqnarray}

The relaxation of the non-magnetic cyclic state to a polar state 
could be faster
than the relaxation to the true ground state, the $F$ state, since 
such relaxation
does not require spin-changing collisions. But on the other hand, the
before-mentioned energy barrier could make the cyclic state quite robust. A
detailed study of the different relaxation rates in a spinor condensates is
certainly warranted.

\section{Superfluid properties}\label{superfluid}

By applying a global rotation ${\cal U}$ together with gauge transformation
$e^{i\theta}$ to the spinor $\hat{\zeta}$ we get a spinor that is physically
identical
\beq
   	\hat{\zeta'}=e^{i\theta}{\cal U}\hat{\zeta}.
\enq
In terms of the Euler angles the rotation operator is given by
\beq
	  {\cal 
U}(\alpha,\beta,\gamma)=e^{iF_z\alpha}e^{iF_y\beta}e^{iF_z\gamma}.
\enq
By making the rotations and gauge transformation local we can induce velocity
fields to the system. Superfluid velocity is defined as ${\bf v}_s=-i
(\hbar/m_a) \hat{\zeta}^\dagger\nabla\hat{\zeta}$ and using this
definition it is straightforward to calculate the superfluid velocities
of the polar, ferromagnetic, and cyclic states:
\begin{eqnarray}
	({\bf v}_s)_{\rm polar}&=&\frac{\hbar}{m_a}\nabla\theta,\\
	({\bf v}_s)_{\rm 
ferro}&=&\frac{2\hbar}{m_a}\left[\nabla(\gamma+\theta/2)+
	\cos\beta\nabla{\alpha}\right],\\
	({\bf v}_s)_{\rm cyclic}&=&\frac{\hbar}{m_a}\nabla\theta.
\end{eqnarray}
The superfluid velocity has the same form for all the polar states and
the ferromagnetic superfluid velocity corresponds to the $F$ state.
Inspecting these results it is clear that the polar and cyclic  states
support only ordinary vortices for which the density at the core is forced to
zero. The ferromagnetic superfluid velocity has the same form as the 
corresponding
result for the $F=1$ spinor condensate~\cite{Ho98}  (there is a factor of $2$
difference, though) and thus coreless vortices are possible although 
they are not expected to be topologically stable.

\section{Discussion}\label{conclusion}

In this paper we have calculated that almost all the excitations of a 
homogeneous
$F=2$ atomic condensate in zero magnetic field are either 
Bogoliubov-type excitations
or free-particle excitations. The only exception is the ferromagnetic $F'$
state, which, however, is never a true ground state. In a strong 
magnetic field all
the excitations have only the Bogoliubov form or the free-particle 
form. We have
also noted that depending on the scattering lengths for collisions 
between atoms in
different $M$ states and on the sign of the quadratic Zeeman coefficient
$g_F^{(2)}$ 
there could be an energy barrier suppressing the decay of the
cyclic state into the (possibly) lower lying polar states.

There are not many physical observables that depend on the relative 
phase of the
different $M$ state wavefunctions in a spinor condensates. We have 
showed that the
value of the chemical potential in the cyclic state depends on the 
existence of relative phase of
the $M=\pm 2,0$ atoms. If there is a well-defined phase relationship 
between these
different components (as there should be, for the cyclic
state), then all physical quantities depending on the chemical 
potential will (in general)  change in a strong magnetic field.

As discussed in the text, for simplicity we have ignored several 
important aspects
of the real physical situation. We have not e.g.~discussed 
dissipative processes or
processes that do not conserve magnetization. Thus our results are 
indicative only,
but we believe they reveal some basic characteristics and important 
{\em a priori}
differences between the $F=1$ and $F=2$ spinor condensates. Regarding the true
experimental situation, the trapping potential has also been absent. 
But since the
main branch of the collective excitations for the cyclic state depends on the
existence of well defined relative phase, 
it can be conjectured that similar dependence should also exist for
the collective excitations of the trapped cyclic state.

\ack
The authors acknowledge the Academy of Finland (project 43336) and the National
Graduate School on Modern Optics and Photonics for financial 
support.

\newpage



\end{document}